\documentstyle[preprint,aps,eqsecnum,version2]{revtex}  
\def\to{\rightarrow}
\def\toffh{e^+ e^- \rightarrow f \bar{f} H}
\def\tozh{e^+ e^- \rightarrow Z H}
\def\toffphi{e^+ e^- \rightarrow f \bar{f} \phi}
\def\tozphi{e^+ e^- \rightarrow Z \phi}
\def\togh{e^+ e^- \rightarrow \gamma H}

\def\simgt{\rlap{\lower 3.5 pt \hbox{$\mathchar \sim$}} \raise 1pt \hbox {$>$}}
\def\simlt{\rlap{\lower 3.5 pt \hbox{$\mathchar \sim$}} \raise 1pt \hbox {$<$}}
\def\epem{\ifmmode{e^+ e^-} \else{$e^+ e^-$} \fi}
\def\Im{\mathop{{\cal I}\mskip-4.5mu \lower.1ex \hbox{\it m}}}
\def\Re{\mathop{{\cal R}\mskip-4mu \lower.1ex \hbox{\it e}}}

\def\gev{{\rm\,GeV}}
\def\tev{{\rm\,TeV}}

\begin{document}
\hfill\parbox{2.5in}{\baselineskip14.5pt
{\bf MAD/PH/734\\
KEK-TH-360\\
KEK preprint 92-222}\\
August 1993}
\vskip.25in
\begin{title}
Probing the Scalar Sector in $\toffh$.
\end{title}
\author{K.~Hagiwara$^1$ \ and M.~L.~Stong$^2$\footnotemark}
\footnotetext{\tenrm Address after 1 Sep.\ 1993: \tenit Inst.\ f\"ur Theor.\
Teilchenphys., Uni. Karlsruhe, Kaiserstr.\ 12 Postfach 6980,
D-76128 Karlsruhe, Germany.}
\begin{instit}  

$^1$Theory Group, KEK, Tsukuba, Ibaraki 305, Japan\\
$^2$Department of Physics, University of Wisconsin, Madison, WI 53706, USA

\end{instit}    
\begin{abstract}
\nonum\section{abstract} 
We study possible deviations from the Standard Model in the reaction
$e^+e^- \to Z\phi$, where $\phi$ denotes a spinless neutral boson.
We show how the $Z$ decay angular correlation can be used to extract
detailed information on the $\phi$ couplings, such as the parity of $\phi$,
radiatively induced form factor effects and possible CP violation in
the scalar sector.
Consequences of gauge invariant dimension six operators are discussed
as an example.
\end{abstract}
\thispagestyle{empty}
\vfil
\newpage
\section{Introduction}

The process $e^+ e^- \to ZH$ is expected to be the best reaction to
look for the Higgs boson ($H$) of mass $\simlt 2m_W$ at LEP II
and at an early stage of the next linear \epem
colliders\cite{lc92,lc500,jlc}.
Unlike in Higgs hunting at hadron colliders\cite{hhunter},
we expect to learn details of the Higgs boson properties and
interactions at \epem colliders.
These include the search for a deviation from the minimal one-doublet
Higgs boson model and for possible radiative effects\cite{zh1,zh2,zh2.5,zh3},
or the effects of the compositeness of $H$ \cite{hllcomp}.
In fact a neutral spinless boson $\phi$ which is produced via
$\epem \to Z\phi$ may not be a Higgs boson at all, but a new
type of a bound state such as a pseudo Nambu-Goldstone boson\cite{ps}
of a new strong interaction with a spontaneously broken chiral
symmetry.
The particle would then be a pseudoscalar rather than a scalar.
We may even expect to observe a CP-violating interaction in the
boson sector\cite{cp1,cp2}.

In this paper, we study couplings of a spinless neutral boson
$\phi$, which may be a scalar, a pseudoscalar, or some mixture of the two,
to the $Z$ boson, in the process $\epem \to Z\phi$; $Z \to f \bar{f}$.
We present general expressions for the production cross sections
and the decay angular distributions with arbitrarily polarized
\epem beams, in terms of the $\toffphi$ helicity amplitudes.
We allow for a general $\phi ZZ$ and $\phi Z\gamma$ vertices
with the mass dimension five that respects the electromagnetic
gauge invariance.
These distributions will then allow us to determine the CP nature of
the $\phi$ boson, and details of the $\phi ZZ$ and $\phi Z\gamma$
interactions.

In the context of the Standard Model, we interpret our generic boson
$\phi$ as a Higgs boson $H$.  Our approach with generic $\phi$ couplings
will be useful in identifying those radiative corrections to the process
$\tozh$\cite{zh1,zh2,zh3} that reduce to the effective $HZZ$ and $HZ\gamma$
form factors.  In the limit of heavy new physics scale, these form factors
can be expressed by the set of gauge-invariant dimension six operators
\cite{bw,hisz} that contribute to these couplings.
Possible CP violation in the Higgs sector may also be observed
as an effective CP-odd vertex of our effective lagrangian.
If $m_H < m_Z$, these couplings may be observed in the decay
$Z \rightarrow \gamma H$ \cite{ah1,ah2,ah3}, and we examine briefly the
consequences of the limits on this process from LEP experiments.

We review here previous studies on related problems.
Kinematics of the process $e^+ e^- \to f\bar{f}H$ have been studied
for the Standard Model at the tree level in \cite{hll1,hll2}.
Beam polarization effects have also been studied in \cite{hllpol}.
Godbole and Roy \cite{hllcomp} introduced the $HZZ$ form factors
in the study of composite light Higgs bosons in the process
$Z \to l\bar{l} H$.  Rattazzi \cite{hll3} has discussed effects of some of
the non-standard couplings that are studied in this paper.  After we
essentially completed this paper, we received a preprint \cite{barger}
in which the $Z$ decay angular distributions in the process $\tozphi$ are
discussed as a means of distinguishing a scalar from a pseudoscalar.
Our results agree with theirs.

The paper is organized as follows.
In section 2 we introduce a phenomenological lagrangian for the
$\phi ZZ$ and $\phi Z\gamma$ couplings, calculate the helicity amplitudes
for the process $\tozphi$, and discuss the information obtainable
using polarized \epem beams.
In section 3 we study the matrix elements for $\toffphi$.  We write the
differential cross section in terms of nine form factors.  These form
factors contain nine combinations of the helicity amplitudes
calculated in section 2.  We form asymmetries which isolate these
combinations, and discuss the measurement properties of those asymmetries.
In section 4 we study effects of the gauge-invariant dimension six
operators in our formalism.  We discuss the limits available from LEP
on the process $\epem \to \gamma H$.  Our conclusions are given in section 5.
\section{Helicity Amplitudes for the Process $\tozphi$.}

We consider an effective lagrangian which contains the Standard Model
couplings of fermions to the $Z$ and $\gamma$, and study the effects
of the following $\phi Z Z$ and $\phi Z\gamma$ couplings:
\begin{eqnarray}
{\cal L}_{ef\! f} = a_Z \, \phi Z^\mu Z_\mu + \sum_V \bigg\{ &&
b_V \, \phi Z^{\mu \nu} V_{\mu \nu} +
c_V \, \left[ (\partial_\mu \phi ) Z_\nu - (\partial_\nu \phi ) Z_\mu \right]
                V^{\mu \nu}  \nonumber \\
&+& \tilde{b}_V \, \phi Z^{\mu \nu} \widetilde{V}_{\mu \nu} \:\:\: \bigg\},
\label{leff}
\end{eqnarray}
where $V_{\mu \nu} = \partial_\mu V_\nu - \partial_\nu V_\mu$, and
$\widetilde{V}_{\mu \nu} \equiv {1 \over 2}
\varepsilon_{\mu \nu \alpha \beta} V^{\alpha \beta}$, with the convention
$\varepsilon_{0 1 2 3} = + 1$.
The terms $a_Z$, $b_V$, and $c_V$ alone would correspond to a CP-even
scalar $\phi$, while the terms $\tilde{b}_V$ alone indicate a CP-odd
pseudoscalar $\phi$.
The presence of both sets of terms tells that $\phi$ is not a CP
eigenstate.  Interference of these two sets of terms
leads to CP-violating effects in the differential cross section.
In the SM, only the coefficient $a_Z$ is non-zero at the tree level,
where $a_Z = g_Z m_Z/2$.
We note here that the terms which are obtained from the $c_V$ terms by
replacing $V^{\mu \nu}$ by $\widetilde{V}^{\mu \nu}$ in the above lagrangian
are equivalent to the negative of the $\tilde{b}_V$ terms.

These effective interactions contribute to the process $\tozphi$
as shown in Fig.~\ref{fig:feyn}.
For electron helicity $\tau = \pm 1$ (in units of $\hbar/2$ \cite{hz})
and outgoing $Z$ boson helicity $\lambda$, the helicity amplitudes
for $\tozphi$ are given by:
\begin{eqnarray}
{ \cal M}_\tau^{(\lambda = 0)} &=& \widehat{M}_\tau^{(\lambda = 0)}
                                        \tau \sin \theta, \nonumber \\
{ \cal M}_\tau^{(\lambda = \pm 1)} &=& \widehat{M}_\tau^{(\lambda = \pm 1)}
                        { 1 + \lambda \tau \cos \theta \over \sqrt{2}},
\label{calm}
\end{eqnarray}
where
\begin{eqnarray}
\widehat{M}_\tau^{(\lambda = 0)} &=&
        { g_Z \left( v_e + \tau a_e \right) \over s - m_Z^2 + i m_Z \Gamma_Z}
        \bigg( 2 \sqrt{s} \, {\omega \over m_Z} ( a_Z + (s+m_Z^2) c_Z )
        + 4 s m_Z (b_Z - c_Z) \bigg)
\nonumber \\ \label{mhat0} &&
        - e \left( 2 \sqrt{s} {\omega \over m_Z}  c_\gamma
        + 2 m_Z (b_\gamma - c_\gamma) \right), \\
\widehat{M}_\tau^{(\lambda = \pm 1)} &=&
        {g_Z \left( v_e + \tau a_e \right) \over s - m_Z^2 +i m_Z \Gamma_Z}
        \bigg( 2 \sqrt{s} ( a_Z + (s + m_Z^2) c_Z )
        + 4 s \omega (b_Z - c_Z) -i \lambda 4 s k \tilde{b}_Z \bigg)
\nonumber \\ \label{mhat1} &&
        - e \left( 2 \sqrt{s} c_\gamma + 2 \omega (b_\gamma - c_\gamma)
        - i \lambda 2 k \tilde{b}_\gamma \right).
\end{eqnarray}
Here $\theta$ is the polar angle of the $Z$ momentum about the electron
beam direction, $\omega = \sqrt{s} \, (1 + m_Z^2/s - m_\phi^2/s)/2$
and $k = (\omega^2 -m_Z^2)^{1/2}$ are the $Z$ energy and momentum
in the $e^+ e^-$ c.m.\ frame.
The couplings are denoted by $e=\sqrt{4\pi\alpha}$,
$g_Z=e/(\sin\theta_W \cos\theta_W)$, and,
$v_e=-1/4+\sin^2\theta_W$ and $a_e=1/4$ are the vector and axial-vector
couplings of the $Z$ to the electron.

These helicity amplitudes may be used directly to produce cross sections
for arbitrarily polarized beams \cite{hz}.
If we denote the transverse polarization directions ${\bf \hat{s}}_\pm$
of $e^\pm$ as
\begin{eqnarray} \label{shat}
{\bf \hat{s}}_\pm = (\cos \varphi_\pm , \sin \varphi_\pm , 0),
\end{eqnarray}
where the azimuthal angles in the $x-y$ plane are measured from the
$x$-axis defined by the outgoing $Z$ transverse momentum,
we can express the $e^\pm$ spin vectors as
\begin{eqnarray} \label{smu}
s_\pm^\mu = P_\pm^T (0,{\bf \hat{s}}_\pm) +
                P_\pm^L (|{\bf p}_\pm|,E_\pm{\bf \hat{p}}_\pm)/m_e.
\end{eqnarray}
The beam polarizations are limited by $0 \le P_\pm^T \le \sqrt{1-(P_\pm^L)^2}$
with $-1 \le P_\pm^L \le 1$.
Purely left-handed $e^\pm$ beams give $P_\pm^L = -1$ and purely right-handed
$e^\pm$ beams give $P_\pm^L = +1$.  Natural transverse polarization of the $e^+
e^-$
storage ring colliders gives $\varphi_+ = \varphi_- +\pi$ and
$P_+^T = P_-^T$.
Arbitrarily polarized beams will be available at $e^+ e^-$ linear colliders.

We can now obtain the matrix element-squared for $\tozphi$, with
arbitrarily polarized $e^+ e^-$ beams, summed over $Z$ polarizations,
by choosing the transverse spin directions as
\begin{eqnarray}
\varphi_- &=& -\varphi,  \nonumber \\ \label{phi.pol}
\varphi_+ &=& -\varphi +\pi +\delta,
\end{eqnarray}
where $\varphi$ is the azimuthal angle of the $Z$ momentum as measured
from the electron transverse momentum direction, and $\pi +\delta$ is
the relative opening angle of the electron and positron transverse
polarizations.  We find
\begin{eqnarray}  \nonumber
\overline{\sum} | {\cal M} |^2 = && {1 \over 4} \sum_\lambda \left\{
        ( 1 + P_-^L)(1 - P_+^L) | {\cal M}_+^\lambda |^2 +
        ( 1 - P_-^L)(1 + P_+^L) | {\cal M}_-^\lambda |^2 \right. \\ \nonumber
& & \qquad{} + \left. 2 P_-^T P_+^T
\left[\cos(2\varphi - \delta)\Re({\cal M}_-^\lambda {\cal M}_+^{\lambda *}) +
\sin(2\varphi - \delta)\Im({\cal M}_-^\lambda {\cal M}_+^{\lambda *}) \right]
\right\} \\ \nonumber
\hfill = && {1 \over 4} \left\{
(1 + P_-^L)(1 - P_+^L) \left[ |\widehat{M}_+^0|^2 \sin^2 \theta
        + |\widehat{M}_+^+|^2 {(1 + \cos \theta)^2 \over 2}
        + |\widehat{M}_+^-|^2 {(1 - \cos \theta)^2 \over 2}
        \right] \right. \\ \nonumber
&&  + (1 - P_-^L)(1 + P_+^L) \left[ |\widehat{M}_-^0|^2 \sin^2 \theta
        + |\widehat{M}_-^+|^2 {(1 - \cos \theta)^2 \over 2}
        + |\widehat{M}_-^-|^2 {(1 + \cos \theta)^2 \over 2}
        \right] \\ \nonumber
&& + 2 P_-^T P_+^T \sin^2 \theta \left[ \cos(2\varphi - \delta)
        \Re \left\{ (\widehat{M}_-^0 \widehat{M}_+^{0 *})
        + {1 \over 2} (\widehat{M}_-^+ \widehat{M}_+^{+ *})
        + {1 \over 2} (\widehat{M}_-^- \widehat{M}_+^{- *}) \right\}
        \right. \\
&& \left. \left. \hphantom{+ 2 P_-^T P_+^T }
+ \sin(2\varphi - \delta)
\Im \left\{ (\widehat{M}_-^0 \widehat{M}_+^{0 *})
        + {1 \over 2} (\widehat{M}_-^+ \widehat{M}_+^{+ *})
        + {1 \over 2} (\widehat{M}_-^- \widehat{M}_+^{- *}) \right\}
        \right] \right\}.    \label{msquard}
\end{eqnarray}
By inserting the helicity amplitudes (\ref{mhat0},~\ref{mhat1}), we find
\begin{eqnarray} \nonumber
\overline{\sum} | {\cal M} |^2  = && \displaystyle {1 \over
        (s- m_Z^2)^2 + m_Z^2 \Gamma_Z^2} \\ \nonumber
        & \hfill \times & \bigg\{ \left(1 - P_-^L P_+^L \right) \bigg[
        2 g_Z^2  (v_e^2 + a_e^2) s \bigg( \sin^2 \theta A_W^2
        + (1 + \cos^2 \theta) \left( A_0^2 + \widetilde{B}_Z^2 \right)
                                \bigg) \\ \nonumber
        &\hfill - & \qquad{}2 g_Z e v_e ( s - m_Z^2)
        \bigg( \sin^2 \theta A_W G_W
        + (1 + \cos^2 \theta ) \left( A_0 G_0 + \widetilde{B}_Z
                \widetilde{B}_\gamma \right) \bigg) \\ \nonumber
        &\hfill - & \qquad{}4 g_Z e a_e m_Z \Gamma_Z \cos \theta
                \left( A_0 \widetilde{B}_\gamma - \widetilde{B}_Z G_0 \right)
        \\ \nonumber
        &\hfill + & \qquad{}e^2 \left( (s - m_Z^2)^2 + m_Z^2 \Gamma_Z^2 \right)
        {1 \over 2 s} \left( \sin^2 \theta G_W^2 + (1 + \cos^2 \theta)
        \left(  G_0^2 +\widetilde{B}_\gamma^2 \right) \right)
        \bigg] \\ \nonumber
        & \hfill + & \left(P_-^L - P_+^L \right) \bigg[
        4 g_Z^2 v_e a_e s
        \bigg( \sin^2 \theta A_W^2
        + (1 + \cos^2 \theta) \left( A_0^2 + \widetilde{B}_Z^2 \right)
        \bigg) \\ \nonumber
        &\hfill - & \qquad{}2 g_Z e a_e (s - m_Z^2)
        \bigg( \sin^2 \theta A_W G_W + (1 + \cos^2 \theta )
        \left( A_0 G_0 + \widetilde{B}_Z \widetilde{B}_\gamma \right)
        \bigg) \\ \nonumber
        &\hfill - & \qquad{}4 g_Z e v_e m_Z \Gamma_Z \cos \theta
        \left( A_0 \widetilde{B}_\gamma - \widetilde{B}_Z G_0 \right)
        \bigg] \\ \nonumber
        &\hfill + & 2 P_-^T P_+^T \sin^2 \theta \bigg[
        g_Z^2  (v_e^2 - a_e^2) s
        \left( A_0^2 - A_W^2 + \widetilde{B}_Z^2 \right) \cos (2\varphi-\delta)
        \\ \nonumber
        & \hfill - & \qquad{}g_Z e v_e (s - m_Z^2)
        \left( A_0 G_0 - A_W G_W + \widetilde{B}_Z \widetilde{B}_\gamma \right)
        \cos (2 \varphi -\delta) \\ \nonumber
        & \hfill + & \qquad{}e^2 \left( (s - m_Z^2)^2 + m_Z^2 \Gamma_Z^2
\right)
        {1 \over 4 s} \left( G_0^2 - G_W^2 +  \widetilde{B}_\gamma^2\right)
        \cos (2 \varphi -\delta) \\
        & \hfill - & \qquad{}g_Z e a_e m_Z \Gamma_Z
        \left( A_0 G_0 - A_W G_W +  \widetilde{B}_Z \widetilde{B}_\gamma\right)
        \sin (2\varphi -\delta)  \bigg] \bigg\}. \label{msqamp}
\end{eqnarray}
Here the following definitions have been used:
\begin{eqnarray}
A_0 &=& a_Z + (s + m_Z^2) c_Z + 2 \sqrt{s} \omega (b_Z - c_Z), \\
A_W &=& {\omega \over m_Z} (a_Z + (s + m_Z^2) c_Z)
                + 2 \sqrt{s} m_Z (b_Z - c_Z), \\
G_0 &=& 2 s c_\gamma + 2 \sqrt{s} \omega (b_\gamma - c_\gamma ), \\
G_W &=& 2 s {\omega \over m_Z} c_\gamma
                + 2 \sqrt{s} m_Z (b_\gamma - c_\gamma), \\
\widetilde{B}_Z &=& 2 \sqrt{s} k \tilde{b}_Z , \\
\widetilde{B}_\gamma &=& 2 \sqrt{s} k \tilde{b}_\gamma ,
\end{eqnarray}
and we assumed that the coefficients in the effective lagrangian (\ref{leff})
do not have imaginary parts.
Only the terms $A_0$ and $A_W$ have a tree-level contribution for the
SM Higgs boson.  The differential cross section is expressed as
\begin{eqnarray} \label{dsig}
{d \sigma \over d \cos \theta d \varphi} = {1 \over 64\pi^2 s}
\overline{\beta}({m_Z^2 \over s},{m_\phi^2 \over s})
\overline{\sum} | {\cal M} |^2
\end{eqnarray}
where $\overline{\beta}(a,b) = (1+a^2+b^2-2a-2b-2ab)^{1/2}$ is the
two-body phase space suppression factor.

We observe the following points.
First, transverse polarization of the beams does not give us new information
even with arbitrary relative opening angle $\pi +\delta$ between the
electron and positron polarizations.
Second, longitudinal polarization is useful for measuring the $HZ\gamma$
couplings in the $G_0$ factor from the interference term $A_0 G_0$,
since the polarization asymmetry of the relevant term is proportional
to $a_e$ whose magnitude is much bigger than that of $v_e$ which
multiplies the $A_0 G_0$ term for $P_-^L=P_+^L$.  Third,
forward-backward asymmetry signals CP violation because the identity
\begin{eqnarray} \label{cp}
|{\cal M}_\tau^{\lambda}(\theta)| =
|{\cal M}_\tau^{-\lambda}(\pi-\theta)|
\end{eqnarray}
holds under CP invariance.  The asymmetry is not only CP-odd but also
CP$\widetilde{\rm T}$-odd \cite{hpzh}, and hence it is proportional to
$\Gamma_Z$ in our approximation of neglecting imaginary parts in the
effective couplings.

The total cross section is simply obtained from the differential cross
section of eq.~(\ref{dsig}).  Fig.~\ref{fig:sm} gives the total cross section
for $\epem \to ZH$ in the Standard Model.  A $300$ GeV linear collider
with a yearly integrated luminosity of $10 {\rm fb}^{-1}$ \cite{jlc}
is expected to produce more than a thousand Higgs bosons in the intermediate
mass range ($m_H \simlt 2 m_W$).  With a good detection efficiency, we can
hope to measure the detailed properties of the Higgs couplings to gauge
bosons.  In Fig.~\ref{fig:xsect}, we show the effect on
the total cross section of adding small couplings $b_Z$, $c_Z$, and
$\tilde{b}_Z$ to the Standard Model. The CP-even couplings $b_Z$ and $c_Z$
have interference terms with $a_Z$, and their
effect is accordingly much larger than that of $\tilde{b}_Z$, which is
CP-odd and appears only quadratically in the total cross section.
The sensitivity of the total cross section to the $\phi Z \gamma$
couplings $b_\gamma$ and $c_\gamma$ is rather poor without beam polarization,
as expected.  In Fig.~\ref{fig:ALR}, we show the polarization asymmetry
\begin{equation}\label{polasy}
A_{\rm LR} = { \sigma_{tot}(P_-^L = -1,P_+^L = 0) -
                        \sigma_{tot}(P_-^L = 1,P_+^L = 0) \over
                \sigma_{tot}(P_-^L = -1,P_+^L = 0) +
                        \sigma_{tot}(P_-^L = 1,P_+^L = 0)}
\end{equation}
as a function of $b_\gamma$, $c_\gamma$, and $\tilde{b}_\gamma$ for
$(\sqrt{s},m_H) = (200,60) \gev$ (a) and $(300,150) \gev$ (b).  When the
$\phi Z \gamma$ couplings are set equal to zero, this asymmetry is
independent of $b_Z$, $c_Z$, and $\tilde{b}_Z$.  This can easily be read
off from eq.~(\ref{msqamp}) by setting $G_0 = G_W = \widetilde{B}_\gamma
= 0$, which leaves exactly the same combination of matrix elements as
coefficients of $(1 - P_-^L P_+^L)$ and $(P_-^L - P_+^L)$.
\section{Decay Angular Correlations}

For electron helicity $\tau$ and outgoing fermion helicity $\tau^\prime$,
the matrix elements $T_{\tau}^{\tau^\prime}$ for $\toffphi$
are given in terms of the helicity amplitudes ${\cal M}_\tau^\lambda$
(\ref{calm}) for $\tozphi$ by
\begin{eqnarray} \label{tamp}
T_{\tau}^{\tau^\prime} (\theta,\hat{\theta},\hat{\varphi})
         = \sum_{\lambda} {\cal M}_\tau^\lambda (\theta )
                {1 \over q^2 - m_Z^2 + i m_Z \Gamma_Z}
                D_\lambda^{\tau^\prime} (q^2,\hat{\theta},\hat{\varphi}),
\end{eqnarray}
where $q^2$ is the invariant-mass squared of the decaying $Z$.
The two quantities ${\cal M}_\tau^\lambda$ and $D_\lambda^{\tau \prime}$
are Lorentz scalars and may therefore
each be evaluated in the reference frame where they are simplest.
${\cal M}_\tau^\lambda$ is evaluated in the $e^+ e^-$ center-of-mass frame,
with the
$Z$ boson momentum in the $z'$-axis.  The angle $\theta$, as before,
separates the $e^-$ and $Z$ momenta which lie in the $x'-z'$ plane.
$D_\lambda^{\tau \prime}$ is evaluated in the
$Z$ rest frame reached by boosting along the $z'$-axis.
The angle $\hat{\theta}$ is then the angle between the $z'$-axis
and the outgoing fermion momentum, and
$\hat{\varphi}$ is the azimuthal angle between the outgoing fermion
and the $x'$-$z'$ plane.
This choice ensures that the $Z$ polarization vectors
\begin{eqnarray}
\varepsilon^\mu_{(\lambda = \pm 1)} &=& \mp(0,1,\pm i,0) / \sqrt{2}
        = (0,-\lambda,-i,0)/ \sqrt{2}, \nonumber \\ \label{epsilon}
\varepsilon^\mu_{(\lambda = 0)} &=& (q,0,0,E)/m_Z,
\end{eqnarray}
contained in ${\cal M}_\tau^\lambda$ and $D_\lambda^{\tau \prime}$
are evaluated in the two frames by proper substitution of the
three-momentum $q$ and energy $E$ of the $Z$.
The decay amplitudes are then
\begin{eqnarray} \label{bigd}
D_\lambda^{\tau^\prime} (q^2,\hat{\theta}, \hat{\varphi})
        = g_Z \sqrt{q^2} \left( v_f + \tau^\prime a_f
                \right) d_\lambda^{\tau^\prime}
                \left( \hat{\theta}, \hat{\varphi} \right),
\end{eqnarray}
where
\begin{eqnarray}
d_{(\lambda = 0)}^{\tau^\prime}
                \left( \hat{\theta}, \hat{\varphi} \right) &=&
            \tau^\prime \sin \hat{\theta}, \nonumber \\ \label{smalld}
d_{(\lambda = \pm 1)}^{\tau^\prime}
                \left( \hat{\theta}, \hat{\varphi} \right) &=&
            {1 \over \sqrt{2}} (1 \pm \tau^\prime \cos \hat{\theta} )
                        e^{\pm i \hat{\varphi}}. \nonumber
\end{eqnarray}

The differential cross section for a given electron helicity $\tau$
and the final fermion helicity $\tau'$,
summed over $Z$ polarizations and integrated over $q^2$, is
\begin{eqnarray} \label{dsig2}
{d \sigma \left( \tau, \tau^\prime \right) \over d \! \cos \! \theta
                \, d \! \cos \! \hat{\theta} \, d \hat{\varphi} } =
        {1 \over 32 \pi s} && \bar{\beta} \left({m_Z^2 \over s},
                                        {m_\phi^2 \over s} \right)
{3 {\rm B}(Z \to f \bar{f}) \over 16 \pi}
{ (v_f + \tau^\prime a_f)^2 \over 2(v_f^2 + a_f^2)} \\ \nonumber
&& \times \left| \sum_\lambda
                {\cal M}_\tau^\lambda (\theta)
                d_\lambda^{\tau^\prime} (\hat{\theta},\hat{\varphi})
         \right|^2,
\end{eqnarray}
where use has been made of the limit $\Gamma_Z \ll m_Z$.
We can expand the squared matrix elements above in terms of the nine
independent decay angular distributions:
\begin{eqnarray}
\left| \sum_\lambda {\cal M}_\tau^\lambda (\theta)
                    d_\lambda^{\tau^\prime} (\hat{\theta},\hat{\varphi})
\right|^2 = &&
        {\cal F}_1 (1 + \cos^2 \hat{\theta}) +
        {\cal F}_2 (1 - 3 \cos^2 \hat{\theta}) +
        {\cal F}_3 \cos \hat{\theta} \nonumber \\ &&
        + {\cal F}_4 \sin \hat{\theta} \cos \hat{\varphi} +
        {\cal F}_5 \sin (2 \hat{\theta}) \cos \hat{\varphi} +
        {\cal F}_6 \sin^2 \hat{\theta} \cos (2 \hat{\varphi}) \nonumber \\ &&
        + {\cal F}_7 \sin \hat{\theta} \sin \hat{\varphi} +
        {\cal F}_8 \sin (2 \hat{\theta}) \sin \hat{\varphi} +
        {\cal F}_9 \sin^2 \hat{\theta} \sin (2 \hat{\varphi}).
\label{fi}
\end{eqnarray}
The angular distributions are defined such that only the coefficient
${\cal F}_1$ remains after integration over the decay angles
$\hat{\theta}$ and $\hat{\varphi}$.

The coefficients ${ \cal F}_i$ are expressed compactly in terms of
the hatted matrix elements in eqs.(\ref{mhat0},~\ref{mhat1}):
\begin{eqnarray}
{ \cal F}_1 &=& {1 + \cos^2 \theta \over 2}
                \left( |\widehat{M}_\tau^+|^2
                + |\widehat{M}_\tau^-|^2 \right)
                + \sin^2 \theta  |\widehat{M}_\tau^0|^2
                + \tau \cos \theta
                \left( |\widehat{M}_\tau^+|^2
                - |\widehat{M}_\tau^-|^2 \right),
                \label{f1} \\
{ \cal F}_2 &=& \sin^2 \theta |\widehat{M}_\tau^0|^2,
                \label{f2} \\
{ \cal F}_3 &=& \tau^\prime \left[ (1 + \cos^2 \theta)
                \left( |\widehat{M}_\tau^+|^2
                - |\widehat{M}_\tau^-|^2 \right)
                + 2 \tau \cos \theta
                \left( |\widehat{M}_\tau^+|^2
                + |\widehat{M}_\tau^-|^2 \right)
                \right], \label{f3} \\
{ \cal F}_4 &=& 2 \tau \tau^\prime \sin \theta \left\{
                \Re \left[ (\widehat{M}_\tau^+ + \widehat{M}_\tau^-)
                        (\widehat{M}_\tau ^0)^\ast \right]
                + \tau \cos \theta
                \Re \left[ (\widehat{M}_\tau^+ - \widehat{M}_\tau^-)
                        (\widehat{M}_\tau ^0)^\ast \right] \right\},
                \label{f4} \\
{ \cal F}_5 &=& \tau \sin \theta \left\{
                \Re \left[ (\widehat{M}_\tau^+ - \widehat{M}_\tau^-)
                        (\widehat{M}_\tau ^0)^\ast \right]
                + \tau \cos \theta
                \Re \left[ (\widehat{M}_\tau^+ + \widehat{M}_\tau^-)
                        (\widehat{M}_\tau ^0)^\ast \right] \right\},
                \label{f5} \\
{ \cal F}_6 &=& \sin^2 \theta
                \Re \left[ (\widehat{M}_\tau^+)(\widehat{M}_\tau^-)^\ast
                \right],
                \label{f6} \\
{ \cal F}_7 &=& - 2 \tau \tau^\prime \sin \theta \left\{
                \Im \left[ (\widehat{M}_\tau^+ - \widehat{M}_\tau^-)
                        (\widehat{M}_\tau ^0)^\ast \right]
                + \tau \cos \theta
                \Im \left[ (\widehat{M}_\tau^+ + \widehat{M}_\tau^-)
                        (\widehat{M}_\tau ^0)^\ast \right] \right\},
                \label{f7} \\
{ \cal F}_8 &=& - \tau \sin \theta \left\{
                \Im \left[ (\widehat{M}_\tau^+ + \widehat{M}_\tau^-)
                        (\widehat{M}_\tau ^0)^\ast \right]
                + \tau \cos \theta
                \Im \left[ (\widehat{M}_\tau^+ - \widehat{M}_\tau^-)
                        (\widehat{M}_\tau ^0)^\ast \right] \right\},
                \label{f8} \\
{ \cal F}_9 &=& - \sin^2 \theta
                \Im \left[ (\widehat{M}_\tau^+)(\widehat{M}_\tau^-)^\ast
                \right].
                \label{f9}
\end{eqnarray}
It is clear that there are nine quantities of interest which we can
obtain from these.  We define primed and unprimed functions which
isolate these nine quantities.
\begin{eqnarray} \label{ai}
f_i(\tau,\tau^\prime) &=&  \int_{-1}^{1} d \cos \theta
        { \cal F}_i (\tau, \tau^\prime),
\\ \label{ai'}
f_i^\prime(\tau, \tau^\prime) &=& \left( \int_0^1 - \int_{-1}^{0} \right)
                d \cos \theta { \cal F}_i(\tau, \tau^\prime),
\end{eqnarray}
and we find:
\begin{eqnarray} \label{a1}
f_1(\tau, \tau^\prime) & = & {4 \over 3} \left[ |\widehat{M}_\tau^+|^2
        + |\widehat{M}_\tau^-|^2 + |\widehat{M}_0|^2 \right], \\ \label{a1'}
f_1^\prime(\tau, \tau^\prime) & = & \tau
                \left( |\widehat{M}_\tau^+|^2
                - |\widehat{M}_\tau^-|^2 \right),
                \\ \label{a2}
f_2(\tau, \tau^\prime) & = & {4 \over 3} |\widehat{M}_\tau^0|^2, \\ \label{a2'}
f_2^\prime(\tau, \tau^\prime) & = & 0, \\ \label{a3}
f_3 (\tau, \tau^\prime)& = & {8 \over 3} \tau^\prime
                \left( |\widehat{M}_\tau^+|^2
                - |\widehat{M}_\tau^-|^2 \right),
                \\ \label{a3'}
f_3^\prime (\tau, \tau^\prime) & = & 2 \tau \tau^\prime
                \left( |\widehat{M}_\tau^+|^2
                + |\widehat{M}_\tau^-|^2 \right),
                \\ \label{a4}
f_4 (\tau, \tau^\prime) & = & \pi \tau \tau^\prime
                \Re \left[ (\widehat{M}_\tau^+ + \widehat{M}_\tau^-)
                        (\widehat{M}_\tau ^0)^\ast \right],
                \\ \label{a4'}
f_4^\prime (\tau, \tau^\prime) & = & {4 \over 3} \tau^\prime
                \Re \left[ (\widehat{M}_\tau^+ - \widehat{M}_\tau^-)
                        (\widehat{M}_\tau ^0)^\ast \right],
                \\ \label{a5}
f_5 (\tau, \tau^\prime) & = & {\pi \over 2} \tau
                \Re \left[ (\widehat{M}_\tau^+ - \widehat{M}_\tau^-)
                        (\widehat{M}_\tau ^0)^\ast \right],
                \\ \label{a5'}
f_5^\prime (\tau, \tau^\prime) & = & {2 \over 3}
                \Re \left[ (\widehat{M}_\tau^+ + \widehat{M}_\tau^-)
                        (\widehat{M}_\tau ^0)^\ast \right],
                \\ \label{a6}
f_6 (\tau, \tau^\prime)& = & {4 \over 3}
                \Re \left[ (\widehat{M}_\tau^+)(\widehat{M}_\tau^-)^\ast
                \right],
                \\ \label{a6'}
f_6^\prime (\tau, \tau^\prime)& = & 0, \\ \label{a7}
f_7 (\tau, \tau^\prime)& = & - \pi \tau \tau^\prime
                \Im \left[ (\widehat{M}_\tau^+ - \widehat{M}_\tau^-)
                        (\widehat{M}_\tau ^0)^\ast \right],
                \\ \label{a7'}
f_7^\prime (\tau, \tau^\prime)& = & - {4 \over 3} \tau^\prime
                \Im \left[ (\widehat{M}_\tau^+ + \widehat{M}_\tau^-)
                        (\widehat{M}_\tau ^0)^\ast \right],
                \\ \label{a8}
f_8 (\tau, \tau^\prime)& = & - {\pi \over 2} \tau
                \Im \left[ (\widehat{M}_\tau^+ + \widehat{M}_\tau^-)
                        (\widehat{M}_\tau ^0)^\ast \right],
                \\ \label{a8'}
f_8^\prime (\tau, \tau^\prime)& = & - {2 \over 3}
                \Im \left[ (\widehat{M}_\tau^+ - \widehat{M}_\tau^-)
                        (\widehat{M}_\tau ^0)^\ast \right],
                \\ \label{a9}
f_9 (\tau, \tau^\prime)& = & - {4 \over 3}
                \Im \left[ (\widehat{M}_\tau^+)(\widehat{M}_\tau^-)^\ast
                \right],
                \\ \label{a9'}
f_9^\prime (\tau, \tau^\prime)& = & 0.
\end{eqnarray}

We present in Table~1 the nine combinations of matrix elements which we
have isolated above.  For each, we indicate by a $+$
the terms even under CP and CP$\widetilde{\rm T}$ \cite{hpzh}, while
a $-$ indicates those which are odd under CP and CP$\widetilde{\rm T}$.
Observation of a CP-odd quantity signals CP-violation, whereas
observation of a CP$\widetilde{\rm T}$-odd quantity signals an absorptive
part in the amplitude.  The asymmetries with which we can measure the
matrix elements are also listed.
The asymmetry $A_i^{(\prime)}$ is obtained from the corresponding function
$f_i^{(\prime)}(\tau, \tau^\prime)$ above after one sums over the
electron polarization $(\tau)$, final fermion polarization $(\tau^\prime)$
and species with an appropriate weight (see below).
We put triangles when we should expect suppression of the signal without
beam polarization or final fermion spin measurements.  The
circles indicate that observation of the asymmetry requires identification of
the charge of the final fermion $f$.  The charge identification
requirement is met only for the leptonic decay modes of the $Z$ and
some fraction of the charm and bottom decay modes.  Final spin measurement is
possible only for the mode $Z \to \tau^+ \tau^-$.  Requirement of the beam
polarization does not pose a problem for linear colliders, but may not be
met at LEP-II.

We define integrated asymmetries:
\begin{equation} \label{aidef}
A_i^{(\prime)}(\tau) = {1 \over N} \sum_f \sum_{\tau ^\prime}
        {3 B(Z \to f \bar{f}) \over 16 \pi }
        {(v_f + \tau^\prime a_f)^2 \over 2 (v_f^2 + a_f^2)}
        f_i^{(\prime)}(\tau,\tau^\prime),
\end{equation}
where
\begin{eqnarray} \label{normdef}
N &=& \sum_\tau A_1(\tau), \\  \label{signorm}
\sigma_{tot} & = & {1 \over 32 \pi s}
\bar{\beta}({m_Z^2 \over s},{m_\phi^2 \over s}) N.
\end{eqnarray}
Finally we sum over electron helicities $\tau$ and obtain:
\begin{eqnarray} \label{ai.final}
A_i^{(\prime)} & = & A_i^{(\prime)}(+) + A_i^{(\prime)}(-), \\
\label{ai.LR}
A_{i {\rm LR}}^{(\prime)} & = & A_i^{(\prime)}(+) - A_i^{(\prime)}(-).
\end{eqnarray}
These are summed over all observable $Z$ decay modes and over the
final fermion spins.
It is remarkable that for each combination of matrix elements, one asymmetry
exists which requires measurement neither of the final fermion spin nor of
its charge.

The asymmetries are shown in Figs.~\ref{fig:asy.60} and \ref{fig:asy.150}
as deviations from the Standard Model values caused by adding small couplings
$b_Z$, $c_Z$ and $\tilde{b}_Z$.
A few of the CP-conserving asymmetries are shown in Fig.~\ref{fig:asy.60}a
for $m_H = 60$ GeV and $\sqrt{s} = 200$ GeV.
As the added couplings here are $b_Z$ and $c_Z$, the Higgs is
a pure scalar in this plot.
Some of the CP-nonconserving asymmetries are shown in Fig.~\ref{fig:asy.60}b
as functions  of $\tilde{b}_Z$ (solid lines) and $\tilde{b}_\gamma$ (dotted
lines).  Figures \ref{fig:asy.150}a,~b show the same asymmetries for
$m_H = 150$ GeV and $\sqrt{s} = 300$ GeV.
The CP${\widetilde{\rm T}}$-odd asymmetries $A_1^\prime$, $A_5$, and $A_8$
can be non-vanishing when the $H Z \gamma$ couplings are added.  However,
in our approximation of neglecting the absorptive parts of the amplitudes,
except for the $Z$ boson width, their magnitudes remain very small, even
with the help of beam polarization.
\section{Dimension six operators -- an example}

One possible manifestation of new physics in the scalar sector is the
appearance of the non-renormalizable effective interactions such as
those in the lagrangian (\ref{leff}).  In the electroweak theory, we expect
that
these new interactions form gauge invariant higher dimensional operators
\cite{bw}.  Constraints due to the electroweak gauge invariance will
then relate new physics contributions to the process $\tozh $ to
those in other reactions.  In this section, we examine the effects
of SU(2) $\times$ U(1) gauge-invariant dimension six operators which
involve the Higgs field and the gauge bosons.

There are eleven such operators ${\cal O}_i$ which are CP-even
\cite{bw}.
Of these eleven, only six contribute to the processes $\tozh $
and $Z \to \gamma H$.
In addition, there are five CP-odd operators, $\widetilde{\cal O}_i$:
\begin{eqnarray} \label{ldim6}
{\cal L}_{ef\! f} = \sum_i {f_i \over \Lambda^2} {\cal O}_i
        + \sum_i {\tilde{f}_i \over \Lambda^2} \widetilde{{\cal O}}_i .
\end{eqnarray}
In the notation of ref.~\cite{hisz}, they are expressed as
\begin{eqnarray} \label{oi}
\begin{array}{rll}
{\cal O}_{WW} =& \Phi^\dagger \hat{W}_{\mu \nu} \hat{W}^{\mu \nu} \Phi,
\hspace{20mm}
\widetilde{{\cal O}}_{WW} &= {1 \over 2} \varepsilon_{\mu \nu \alpha \beta}
                \Phi^\dagger \hat{W}^{\mu \nu}
                \hat{W}^{\alpha \beta} \Phi, \\
{\cal O}_{BB} =& \Phi^\dagger \hat{B}_{\mu \nu}
        \hat{B}^{\mu \nu} \Phi, \hfill
\widetilde{{\cal O}}_{BB} &= {1 \over 2} \varepsilon_{\mu \nu \alpha \beta}
                \Phi^\dagger \hat{B}^{\mu \nu}
                \hat{B}^{\alpha \beta} \Phi, \\
{\cal O}_{BW} =& \Phi^\dagger \hat{B}_{\mu \nu}
                \hat{W}^{\mu \nu} \Phi, \hfill
\widetilde{{\cal O}}_{BW} &= {1 \over 2} \varepsilon_{\mu \nu \alpha \beta}
                \Phi^\dagger \hat{B}^{\mu \nu}
                \hat{W}^{\alpha \beta} \Phi,  \\
{\cal O}_W =& (D_\mu \Phi)^\dagger \hat{W}^{\mu \nu}
                (D_\nu \Phi), \hfill
\widetilde{{\cal O}}_W &= {1 \over 2} \varepsilon_{\mu \nu \alpha \beta}
                (D^\mu \Phi)^\dagger \hat{W}^{\alpha \beta}
                (D^\nu \Phi), \\
{\cal O}_B =& (D_\mu \Phi)^\dagger \hat{B}^{\mu \nu}
                (D_\nu \Phi), \hfill
\widetilde{{\cal O}}_B &= {1 \over 2} \varepsilon_{\mu \nu \alpha \beta}
                (D^\mu \Phi)^\dagger \hat{B}^{\alpha \beta}
                (D^\nu \Phi),  \\
{\cal O}_{\Phi, 1} =& (D_\mu \Phi)^\dagger \Phi \Phi^\dagger (D^\mu \Phi),
\hfill & \hfill
\end{array}
\end{eqnarray}
Here the covariant derivative is
\begin{eqnarray} \label{del}
D_{\mu} & = & \partial_\mu + igT^aW^a_\mu + ig'YB_\mu \label{del1} \\
& = & \partial_\mu + i\frac{g}{\sqrt{2}}(W^+_\mu T^+ + W^-_\mu T^-)
+ig_Z(T^3-s_W^2Q)Z_\mu + ieQA_\mu ,
\end{eqnarray}
where $g$ and $g'$ are the SU(2) and U(1) gauge couplings, respectively, and
$c_W=\cos\theta_W$ and $s_W=\sin\theta_W$ are the weak mixing factors
\begin{eqnarray} \label{w3}
\left( \begin{array}{c}
W^3_\mu\\B_\mu \end{array}\right) = \left(\begin{array}{cc}c_W & s_W \\
-s_W & c_W \end{array} \right) \left( \begin{array}{c}
Z_\mu\\A_\mu \end{array} \right). \label{mix}
\end{eqnarray}
The hatted operators are
\begin{eqnarray} \label{what}
\hat{W}_{\mu\nu}  &=&  igT^aW^a_{\mu\nu}, \\ \label{bhat}
\hat{B}_{\mu\nu}  &=&  ig'YB_{\mu\nu},
\end{eqnarray}
and the unhatted field operators are
\begin{eqnarray} \label{wmunu}
\left\{ \begin{array}{l}
W^\pm_{\mu\nu} = \partial_\mu W^\pm_\nu - \partial_\nu W^\pm_\mu \pm ig
(W^3_\mu W^\pm_\nu - W^3_\nu W^\pm_\mu), \\
W^3_{\mu\nu} = \partial_\mu W^3_\nu - \partial_\nu W^3_\mu + ig(W^+_\mu W^-_\nu
-W^+_\nu W^-_\mu), \end{array} \right.
\end{eqnarray}
\begin{eqnarray} \label{bmunu}
B_{\mu\nu} = \partial_\mu B_\nu-\partial_\nu B_\mu.
\end{eqnarray}
The standard Higgs field $\Phi$ is a doublet with the hypercharge
$Y=\frac{1}{2}$, which has the form
\begin{eqnarray} \label{phi.higgs}
\Phi = \left(
\begin{array}{c} 0 \\
\displaystyle { v+H \over \sqrt{2}} \end{array} \right),
\end{eqnarray}
in the unitary gauge.

We note that in the gauge boson sector that has been studied extensively
\cite{gw,hisz} the operators ${\cal O}_{WW}$ and ${\cal O}_{BB}$ merely
renormalize the SM gauge couplings, while the operators ${\cal O}_{BW}$
and ${\cal O}_{\Phi,1}$ contribute to the gauge boson propagators.
All the eleven operators listed above contribute to the $HZZ$ and $HZ\gamma$
couplings and their effects can be expressed as their contributions
to the seven coefficients of our phenomenological lagrangian (\ref{leff}).
They can be calculated simply by counting vertices in ${\cal L}_eff$
\ref{ldim6}, except for the coefficient $a_Z$ (and $b_\gamma = c_\gamma$;
see below) for which we should also count the shifts due to the new
interactions

in the $Z$ and $H$ fields as well as in the paramters $g_Z$ and $m_Z$.

We find by a straightforward calculation
\begin{eqnarray} \label{az}
a_Z &=& {g_Z m_Z \over 2} \left( 1 + f_{\Phi, 1}
                { v^2 \over \Lambda^2} \right), \\ \label{bz}
b_Z &=& - {g_Z m_Z \over 2 \Lambda^2} \left( c_W^4 f_{WW} + s_W^2 c_W^2 f_{BW}
                + s_W^4 f_{BB}\right), \\ \label{cz}
c_Z &=& - {g_Z m_Z \over 4 \Lambda^2}
                \left( c_W^2 f_W + s_W^2 f_B \right), \\ \label{tbz}
\tilde{b}_Z &=& {g_Z m_Z \over 2 \Lambda^2} \left( c_W^4 \tilde{f}_{WW}
                + s_W^2 c_W^2 \tilde{f}_{BW} + s_W^4 \tilde{f}_{BB}
                - {1 \over 2} ( c_W^2 \tilde{f}_W + s_W^2 \tilde{f}_B ),
        \right), \\ \label{ba}
b_\gamma &=& - {e m_Z \over 2 \Lambda^2} \left( 2 c_W^2 f_{WW}
                + (s_W^2- c_W^2) f_{BW}
                - 2 s_W^2 f_{BB}\right), \\ \label{ca}
c_\gamma &=& - {e m_Z \over 4 \Lambda^2}
                \left( f_W - f_B \right), \\ \label{tba}
\tilde{b}_\gamma &=& {e m_Z \over 2 \Lambda^2} \left( 2 c_W^2 \tilde{f}_{WW}
                + (s_W^2- c_W^2) \tilde{f}_{BW} - 2 s_W^2 \tilde{f}_{BB}
                - {1 \over 2} ( \tilde{f}_W - \tilde{f}_B )
                \right), \\ \label{baa}
b_{\gamma \gamma} &=& {e c_W s_W m_Z \over 2 \Lambda^2 }
                \left( f_{WW} + f_{BW}+ f_{BB} \right), \\ \label{tbaa}
\tilde{b}_{\gamma \gamma} &=& - {e c_W s_W m_Z \over 2 \Lambda^2 }
 \left( \tilde{f}_{WW} + \tilde{f}_{BW}+ \tilde{f}_{BB} \right).
\end{eqnarray}
Here we have added to our effective lagrangian (\ref{leff})
the $\phi \gamma \gamma$ couplings:
\begin{equation}
{\cal L}_{eff}(\phi \gamma \gamma) = b_{\gamma \gamma}
                \phi A^{\mu \nu} A_{\mu \nu} + \tilde{b}_{\gamma \gamma}
                \phi A^{\mu \nu} \widetilde{A}_{\mu \nu},
\end{equation}
whose effects can be studied at $\gamma \gamma$ colliders \cite{cp2}
or in the decay $H \to \gamma \gamma$ \cite{hggdecay}.  The dimension
six operators contribute not only to the above $HZZ$, $HZ\gamma$ and
$H\gamma \gamma$ vertices but also to the gauge boson \cite{gw} and the
Higgs propagators.
They shift the boson masses and the wavefunction normalization.
We can be express the result of this renormalization compactly by using the
$Z$ mass and the couplings ($g_Z$ and $s_W^2$) as observed at LEP.
The net effect is then to add to the above effective couplings the following
terms:
\begin{eqnarray}\label{delaz}
\Delta a_Z & = & {g_Z m_Z \over 2}
\left( {f_{\Phi,2} \over 2} {v^2 \over \Lambda^2} - g_Z^2
(c_W^4 f_{DW} + s_W^4 f_{DB}) {2 q^2 -  m_Z^2 \over \Lambda^2} \right),
\\ \label{delbg}
\Delta b_\gamma & = & \Delta c_\gamma = - e g_Z^2
(c_W^2 f_{DW} - s_W^2 f_{DB}) {m_Z \over \Lambda^2},
\end{eqnarray}
in the notation of ref.~\cite{hisz}.  The coefficients $f_{DW}$ and
$f_{DB}$, along with $f_{BW}$ and $f_{\Phi,1}$ are constrained stringently
by the present electroweak measurements \cite{gw,hisz}.  $f_{B}$, $f_{W}$,
$f_{BB}$, and $f_{WW}$ are only mildly constrained at the one-loop level
\cite{hisz} and the remaining CP-even coefficient $f_{\Phi,2}$ is
unconstrained by the low energy data.  All of the CP-odd operators should be
constrained by the non-observation of the neutron and the electron electric
dipole moments \cite{oddcons,BHKN}.  However, a comprehensive analysis of all
the CP-odd dimension six operators has not yet been carried out.  Since only
certain combinations of the five CP-odd operators are constrained by
low-energy experiments, we cannot generally exclude observable CP violation
in the reaction $\toffh$.

In this report, we give just one example of the possible effects of these
new interactions.
The $H Z \gamma$ and $H \gamma \gamma$ couplings above make it possible
for the reaction $\togh$ to occur at a significantly enhanced rate.
The cross section for this process at the $Z$ peak is
\begin{eqnarray} \label{ah}
\sigma (e^+ e^- \rightarrow Z \rightarrow \gamma H ) =
{G_F^2 m_Z^8 \over 3 \pi \Gamma_Z^2}
\left( 1 - {m_H^2 \over m_Z^2} \right)^3 c_W^2 s_W^2
(v_e^2 + a_e^2) \left( {\cal G}^2(f_i) + {\cal G}^2(\tilde{f}_i) \right),
\label{siggh}
\end{eqnarray}
where
\begin{eqnarray} \label{gfi}
{\cal G}(f_i) = ( 2 c_W^2 f_{WW} + (s_W^2 - c_W^2) f_{BW}
                - 2 s_W^2 f_{BB})/ \Lambda^2  - {1 \over 2}
                (f_W  - f_B )/ \Lambda^2 .
\end{eqnarray}
Limits on this process are available from the LEP experiments.
The strictest of these are of the order $1$--$10$ pb \cite{ahlep}.
 From $\sigma(\togh) < 1$ pb, we can obtain a limit of
$\left[{\cal G}^2(f_i) + {\cal G}^2(\tilde{f}_i)\right]
(1 - (m_H^2/m_Z^2))^3 < 20.5 \tev^{-4}$.
This is a limit on $(b_\gamma - c_\gamma) m_Z$ and
on $\tilde{b}_\gamma m_Z$ of $0.022$ for $m_H$ = 70 GeV.
In Figure \ref{fig:photon} we compare the SM contributions to the cross
section at $\sqrt{s} = m_Z $\cite{ah1,ah2,ah3} as the dotted lines,
which are far below the experimental limit (1--10 pb), to the
contribution from the effective couplings $f_{WW}$ (solid line) and
$f_{BB}$ (dashed line).  Also shown in the figure are the effects of these
couplings at higher energies, $\sqrt{s} = 200$ and $300$ GeV.
\section{Conclusions}

We have shown how the $Z$ decay angular correlation in the process
$\toffphi$ is useful in obtaining detailed information on the
CP nature of a spinless neutral particle $\phi$.  To do this, we
have examined the consequences of $\phi ZZ$ and $\phi Z\gamma$ couplings
introduced in an effective lagrangian which includes all terms of mass
dimension five which respect electromagnetic gauge invariance.
For the process $\tozphi$, we have shown that transverse
polarizations of the beams does not give us new information.  Longitudinal
polarization of the beams is useful for studying the CP-even $\phi Z\gamma$
couplings.

Including the decay of the $Z$ provides more information.
There are then nine combinations of the $\tozphi$ matrix elements
which appear in the differential cross section.  One of these combinations
is just the $\tozphi$ cross section summed over $Z$ polarizations.  The
remaining eight are isolated by constructing asymmetries.  For each
combination of matrix elements, we find one asymmetry which isolates it
without identifying the charge or the spin of the final state fermion.
This holds true even for the CP-violating asymmetries.  Therefore
all the observable $Z$ decay modes can be used to measure the $\phi ZZ$
and $\phi Z \gamma$ couplings, without losing generality.  These
results are summarized in Table 1 and also in Figures \ref{fig:asy.60}
and \ref{fig:asy.150}, where we
show the dependence of the asymmetries on the couplings in our effective
lagrangian.

Finally, we have related our effective lagrangian to the special case
of addition of the dimension six operators made of the gauge bosons and
the Higgs doublet to the Standard Model.  Experimental
limits on the $H Z \gamma$ couplings from non-observation of the
process $\togh$ at LEP can be expressed as a constraint on the dimension
six operators when $m_H < m_Z$.
We compare in Figure \ref{fig:photon} the cross section for this
process
from the effective lagrangian to that of the Standard Model at one loop.
\acknowledgements

The authors wish to thank Peter Zerwas for discussions that prompted us to
investigate the problem of measuring the 'Higgs' boson couplings.
One of us (MLS) would also like to thank the NSF's Summer Institute in
Japan program. This work was funded in part by the University of
Wisconsin Research Committee with funds granted by the
Wisconsin Alumni Research Foundation, and in part by the U.S.~Department
of Energy under Contract No.~DE-AC02-76ER00881.

\def\pl #1 #2 #3 {Phys.~Lett. {\bf#1}, #2 (#3)}
\def\np #1 #2 #3 {Nucl.~Phys. {\bf#1}, #2 (#3)}
\def\zp #1 #2 #3 {Z.~Phys. {\bf#1}, #2 (#3)}
\def\pr #1 #2 #3 {Phys.~Rev. {\bf#1}, #2 (#3)}
\def\prep #1 #2 #3 {Phys.~Rep. {\bf#1}, #2 (#3)}
\def\prl #1 #2 #3 {Phys.~Rev.~Lett. {\bf#1}, #2 (#3)}
\def\mpl #1 #2 #3 {Mod.~Phys.~Lett. {\bf#1}, #2 (#3)}
\def\xx #1 #2 #3 {{\bf#1}, #2 (#3)}
\def\etal {{\it et al}.}
\def\eg {{\it e.g}.}
\def\ie {{\it i.e}.}
\def\ibid{{\it ibid}. }

\newpage
\begin{center}
\large TABLE 1
\end{center}
\vspace*{1cm}

CP and CP$\widetilde{\rm T}$ \cite{hpzh} properties of squared matrix elements,
and the corresponding observable asymmetries.
CP and CP$\widetilde{\rm T}$ conservation is indicated with a $+$,
nonconservation is indicated with a $-$.
The circles indicate that observation of the asymmetry requires
identification of the charge of the final fermion $f$. The triangles indicate
that the asymmetry may be suppressed without corresponding polarization
measurements.

\begin{center}
\begin{tabular}{l|cc|c|cc|c} \hline
\multicolumn{1}{c|}{Matrix Elements} &
\multicolumn{2}{c|}{Properties}&
\multicolumn{1}{c|}{~Observables}&
\multicolumn{1}{c}{~beam}&
\multicolumn{1}{c|}{$f$}&
\multicolumn{1}{c}{$f$} \\
 & CP & CP$\widetilde{\rm T}$ & & Pol. & Pol. & charge \\ \hline
$ |\widehat{M}_\sigma^+|^2 + |\widehat{M}_\sigma^-|^2
+ |\widehat{M}_\sigma^0|^2 $ & $+$ & $+$ & $\sigma_{tot}$ & - & - & - \\ \hline
$ |\widehat{M}_\sigma^0|^2 $ & $+$ & $+$ & $A_2$ & - & - & - \\ \hline
$ |\widehat{M}_\sigma^+|^2 - |\widehat{M}_\sigma^-|^2 $ &
$-$ & $-$ & $A_1^\prime$ & $ \bigtriangleup$ & - & - \\ \cline{4-7}
 & & & $A_3$ & - & $\bigtriangleup$ & $\bigcirc$ \\ \hline
$ \Re \left[ (\widehat{M}_\sigma^+ + \widehat{M}_\sigma^-)
(\widehat{M}_\sigma ^0)^\ast \right]$ & $+$ & $+$ & $A_4$ &
$\bigtriangleup$ & $\bigtriangleup$ & $\bigcirc$ \\ \cline{4-7}
 & & & $A_5^\prime$ & - & - & - \\ \hline
$ \Re\left[ (\widehat{M}_\sigma^+ - \widehat{M}_\sigma^-)
(\widehat{M}_\sigma ^0)^\ast \right]$& $-$ & $-$ &
$A_4^\prime$ & - & $\bigtriangleup$ & $\bigcirc$ \\ \cline{4-7}
 & & & $A_5$ & $\bigtriangleup$ & - & - \\ \hline
$\Re \left[ (\widehat{M}_\sigma^+)(\widehat{M}_\sigma^-)^\ast \right]$ &
$+$ & $+$ & $A_6$ & - & - & - \\ \hline
$ \Im\left[ (\widehat{M}_\sigma^+ - \widehat{M}_\sigma^-)
(\widehat{M}_\sigma ^0)^\ast \right]$& $-$ & $+$ &
$A_7$ & $\bigtriangleup$ & $\bigtriangleup$ & $\bigcirc$ \\ \cline{4-7}
 & & & $A_8^\prime$ & - & - & - \\ \hline
$ \Im \left[ (\widehat{M}_\sigma^+ + \widehat{M}_\sigma^-)
(\widehat{M}_\sigma ^0)^\ast \right]$& $+$ & $-$ &
$A_7^\prime$ & - & $\bigtriangleup$ & $\bigcirc$ \\ \cline{4-7}
 & & & $A_8$ & $\bigtriangleup$ & - & - \\ \hline
$\Im \left[ (\widehat{M}_\sigma^+)(\widehat{M}_\sigma^-)^\ast \right]$ &
$-$ & $+$ & $A_9$ & - & - & - \\ \hline
\end{tabular}
\end{center}
\figure{Feynman diagrams for $e^+ e^- \rightarrow Z \phi$, (a) in the
Standard Model, and (b) with our effective Lagrangian.  \
(c) Those for $e^+ e^- \rightarrow \gamma \phi$.\label{fig:feyn}}
\figure{Standard Model cross section for $e^+ e^- \rightarrow Z H$,
for $\sqrt{s} =$ 200, 250, 300, 350, 400, 450, 500 GeV, as a function
of $m_H$.\label{fig:sm}}
\figure{Cross section for $e^+ e^- \rightarrow Z H$, for
($\sqrt{s}$,$m_H$) = (200,60) and (300,150) GeV.  The horizontal solid lines
give the Standard Model values and the solid curves show the dependence on
$b_Z$.  The dashed lines are for $c_Z$, and the dots for
$\tilde{b}_Z$. \label{fig:xsect}}
%
\figure{ $A_{\rm LR}$ in the Standard Model and as a function
of $b_\gamma$, $c_\gamma$, and $\tilde{b}_\gamma$.  When the $HZ\gamma$
couplings are set to zero, this asymmetry is independent of the couplings
$b_Z$, $c_Z$, and $\tilde{b}_Z$. (a) ($\sqrt{s}$,$m_H$) = (200,60) GeV,
and (b) ($\sqrt{s}$,$m_H$) = (300,150) GeV. \label{fig:ALR}}
\figure{Asymmetries for $m_H = 60$ GeV and $\sqrt{s} = 200$ GeV.
(a) $A_2$, $A_5^\prime$, and $A_6$ exist in the Standard Model.
The solid lines indicate the effect of $b_Z$, and the dashed lines
give the effect of $c_Z$.
(b) $A_8^\prime$ and $A_9$ indicate CP violation.
The solid lines give the effect of $\tilde{b}_Z$, the dotted lines of
$\tilde{b}_\gamma$. \label{fig:asy.60}}
\figure{As Fig.~\ref{fig:asy.60}, for $m_H = 150$ and $\sqrt{s} = 300$ GeV.
\label{fig:asy.150}}
\figure{Cross section for $e^+ e^- \rightarrow \gamma H$.
The contribution of $f_{WW}$ ($f_{BB}$) is given by the solid (dashed) lines
for ($\sqrt{s}$,$m_H$) = ($m_Z$,70),(200,60), and (300,150) GeV.
The contribution of the Standard Model (dots) is shown for $m_t = 150$ GeV
and ($\sqrt{s}$,$m_H$) = ($m_Z$,60) GeV.
\label{fig:photon}}
\end{document}